# Bring-Your-Own-Device (BYOD): An Evaluation of Associated Risks to Corporate Information Security

*Ezer Osei Yeboah-Boateng, Ph.D*
Dr. Ezer Osei Yeboah-Boateng is a Senior Lecturer at the Ghana Technology University College (GTUC)

*Francis Edmund Boaten*
Francis Edmund Boaten is the CTO of DiscoveryTel Ghana and holds a Masters in Telecommunications

## ABSTRACT

*This study evaluates the cyber-risks to Business Information Assets posed by the adoption of Bring-Your-Own-Device (BYOD) to the workplace. BYOD is an emerging trend where employees bring and use personal computing devices on the company's network to access applications and sensitive data like emails, calendar and scheduling applications, documents, etc. Employees are captivated by BYOD because they can have access to private items as well as perform certain job functions while being unrestricted to their desks. This is however usually done on the blind side of management or the system administrator; a situation that tends to expose vital and sensitive corporate information to various threats like unwanted network traffic, unknown applications, malwares, and viruses. Expert opinions were elicited in this exploratory study. The study evaluated the characteristics of BYOD, assessed associated risks, threats and vulnerabilities. The findings indicate that little or no security measures were instituted to mitigate risks associated with BYOD.Though, profound benefits abound with BYOD adoption, they could be eroded by security threats and costs of mitigation in curing breaches. The most significant risk was found to be Data Loss which was in consonance with similar studies on Smartphone security risks. Some mitigation measures are then recommended.*

## 1. INTRODUCTION

The use of the Internet has assumed ubiquity, affecting the way we live, work or play. Similarly, the emerging concept of Bring-Your-Own-Device (BYOD), though, relatively new paradigm has caught on amongst many and has assumed tele-commuting purposes. This has been exacerbated by the proliferation of mobile and/or portable devices, such as smartphones, tablets, iPads, notebooks, as well as laptops.

In this study we adopted a similar mobile or portable device definition from(Yeboah-Boateng & Amanor, Phishing, SMiShing & Vishing: An Assessment of Threats against Mobile Devices, 2014) and heretofore call them smartphones.

Another contributing factor is the relative low cost of the smartphone and subsequently, its ease of use at the workplace for both voice and data services.  The smartphone is characterized by its contemporary advanced features and the numerous applications (Apps) that are populated with it.  From the perspective of computer architecture, it's basically an embedded portable computer, equipped with interactive, mostly Java-based Apps, running on flexible operating systems (OSs) such as Android, iOS, RIM, Windows Mobile, to name a few.  The paradigm of running these smartphones on mostly free-and-open-source software (FOSS), have made its adoption and use very plausible and versatile (Hogben & Dekker, 2010).

The key drivers of BYOD adoption and use have been, of course, the proliferation of smartphones, the integral usage of both personal and corporate-provided devices, as well as the need to stay in touch, keep abreast with corporate developments "as-you-go", and tele-commuting for work purposes. Other factors such as, keen competition and the need to meet the expectations of tech-savvy, less-forgiving customers(Yeboah-Boateng, Cyber-Security Challenges with SMEs in Developong Economies: Issues with Confidentiality, Integrity & Availability (CIA), 2013) have culminated into increased productivity. BYOD offers the needed ubiquity – thus, embracing the anywhere, anytime approach to business. Accordingly, organizations are coming to terms with the need for a BYOD strategy in order to authenticate and authorize employees to use BYOD on enterprise networks, which would inure to the benefits of the organization.

Inspite of the above, BYOD presents network externalities which could impact on the corporate security framework of re-defining the uncertainties of network perimeter and to safeguard the information assets to ensure confidentiality, integrity and availability (CIA) (Yeboah-Boateng, Cyber-Security Challenges with SMEs in Developong Economies: Issues with Confidentiality, Integrity & Availability (CIA), 2013)(Ernst & Young, 2013).

## 1.1 Problem Formulation

A key concern with BYOD is invariably it and/or management of the organization may not be aware of personally-owned devices accessing corporate resources. Where any of these are in the known, the needed technical support may not also be forth coming or provided. Studies show that there is an increased cyber-risk posed to the sensitive corporate information assets of any business when "foreign" and/or unauthorized devices access the corporate network (Yeboah-Boateng, Using Fuzzy Cognitive Maps (FCMs) To Evaluate The Vulnerabilities With ICT Assets Disposal Policies, 2012). A compromised mobile device with access to the corporate network could serves as susceptible entry points for nefarious activities within the network and possibly with access to sensitive information. For any organization therefore to adopt BYOD, it is imperative that appropriate security measures be put in place to successfully mitigate against the negative effects of the phenomenon.

In a related development, mobile infections are said to be growing at a disturbing rate, with a whooping increase of 17% in the first half of 2014, as compared with a rate of 20% for the entire 2013 (Aittokallio, 2014). Similarly, mobile-related cyber-attacks is reported to cost UK businesses £18 billion in lost revenue and £16 billion in increased IT expenditure per year resulting from breaches (CEBR, 2015). The report apparently indicates that the mobile devices compromises are a widespread menace, with 81% of UK businesses recording a breach in 2014 (CEBR, 2015). In fact, experts believe that if nothing done about the cyber-risk menace, the

Throughout the world, organizations are faced with cybersecurity breaches mostly resulting in breach costs (e.g., incident response forensics, clean-up, legal), reputation and brand damage, and lost revenue due to downtime and so on (Yeboah-Boateng, Cyber-Security Challenges with SMEs in Developong Economies: Issues with Confidentiality, Integrity & Availability (CIA), 2013). To exacerbate the already precarious situation, most SMEs in developing economies consider themselves as not having any data attractive to threat agents, and are free from attacks, however, the contrary is true as most companies usually have data on employees, customers, suppliers, partners, and even intellectual properties which have the potential of attracting attacks(Yeboah-Boateng, Cyber-Security Challenges with SMEs in Developong Economies: Issues with Confidentiality, Integrity & Availability (CIA), 2013).

Studies have clearly shown that the problem posed by BYOD is rooted in ignorance, and therefore promoting education on security issues associated with BYOD would immensely contribute in addressing the risks and help mitigate the negative effects. The key issues of concern in this study are;

1. What are the Risks, Threats, and Vulnerabilities associated with the adoption of BYODs as working tools in the workplace?
2. What are the most significant risks that organizations in developing economies are exposed to, as a result of BYOD adoption?
3. What mitigation measures can be adopted to address the risks associated with BYOD?
4. What are the characteristics of BYOD use in the workplace?

The study sought to assess the Risks, Threats and Vulnerabilities associated with the adoption of BYODs in the workplace, and to evaluate the most significant Risk that businesses are exposed to, as a result of BYOD adoption. Also, we explore some mitigation measures available to address the associated risks, and to evaluate some characteristics of BYOD use in the workplace.

This introductory section dealt with the overview of BYOD and its concerns in adoption and use on corporate networks. The ensuing sections are the literature review, followed by the research methodology, and then the results and analysis. We conclude with some discussions and recommendations.

## 2. LITERATURE REVIEW

This section reviews the concept of Bring Your Own Device (BYOD) and its associated risks to the corporate information assets. It then proceeds to assess specific Risks, Threats, and Vulnerabilities, associated with BYOD adoption and use, and also explains strategies that can be adopted for their mitigation.

### 2.1 Bring Your Own Device (BYOD)

Bring Your Own Device (BYOD) is a new development in the workplace where employees are encouraged to access company resources like corporate e-mails, calendars and scheduling, documents, applications and so on with their personal devices, either for work or for personal use (Ghosh, Gajar, & Rai, 2013).

BYODs are usually portable mobile computing devices, and are used for numerous computing applications and services. For purposes of this study, BYODs include Smartphones, Laptops, Notebooks, iPads, Tablets, etc. and adopted the term "smartphone" for these BYOD devices. These devices could run on any operating system, like Android, iOS, or RIM and possess peculiar features like their small physical size and light weight, advanced computing features, including processing power and storage, making them extremely useful and versatile. They likewise have the capability to connect from anywhere via hotspots anytime within and outside the workplace.

In the early 2000s when the first BlackBerry was out-doored, industries saw the paybacks of remote email and calendar access and began providing smartphones which came with network access to many employees, successfully establishing the idea of anywhere anytime connectivity.

Intel, in 2009, first acknowledged the movement towards BYOD, when employees progressively sought to use their personal mobile devices in the office. The management made a good judgement and instead of overlooking the potential risk, they accepted the technology, and came up with an operational policy for personal devices. The result was improved connectivity to Intel's network, greater worker throughput and enhanced security procedures (Field, 2011).

The endorsement of smartphones went beyond business users when Apple's iPhone was released. Later Android and Windows Mobile Phone 7 also came along. Features extended beyond traditional email and web browsing, and now devices have the capability of photography, running customized applications, viewing rich content websites with Flash and JavaScript, establishing Wi-Fi connections, establishing virtual private network (VPN) connections, as well as acting as tethering or hotspots(Belletati, 2014).

## 2.2   BYOD Associated Vulnerabilities
The ensuing sub-sections review literature on vulnerabilities presented by BYOD use. Generally, vulnerabilities are weaknesses or flaws inherent in a computer system, which make corporate information and applications susceptible, and consequently are exploited by threat agents.

**Vulnerabilities facilitating Malware**
Malware is one of the leading threats to mobile devices (Yeboah-Boateng & Amanor, Phishing, SMiShing & Vishing: An Assessment of Threats against Mobile Devices, 2014)therefore any vulnerability leading to or making it plausible for a malware to be installed on a mobile device needs to be addressed. Malware include viruses, Trojans, spyware, financial malware, ransomware and so on.

**User Permissions Issues**
Most platforms require a user's permission for applications to access various types of data on the device at the time of installation of the application. It is often very cumbersome to review and change the permissions users have granted at the time of installation and also there is no means to set global policies for permissions granted, for instance to generally instruct that no apps should be installed requesting location information for marketing purposes(Ernst & Young, 2012).

**Encryption Issues**
There are conspicuous weaknesses in some implementations of mobile device encryption(Whitman & Mattord, Management of Information Security, 2004), if any at all, which makes data protection on these devices extremely porous. Attackers can capitalize on this weakness once they have physical access to a lost or stolen device or when under repairs. The effectiveness of encryption schemes largely depends on the technical procedures and measures used for managing cryptographic keys(Nunoo, 2013).

**User Awareness Issues**
The lack of user awareness or user naivety can be blamed for unintentional disclosure of data and other vulnerabilities, as for instance most users are completely ignorant of the implications of consenting to certain kinds of data disclosure through the installation process as we have discussed earlier.

## 2.3   BYOD related Threats & Associated Risks
The following sub-sections review literature on threats and risks presented by BYOD use.

**Data Leakage due to Lost or Stolen Devices**
Data Leakage can be the result of an attacker accessing data on a misplaced or stolen mobile device with unprotected memory. BYODs are usually in high demand and handy, and are likely to be stolen or misplaced or undergo repairs. If data on the BYOD memory or its removable media is not adequately secured by encryption, an attacker can access that information(Causey, 2013). BYODs often contain valuable information like credit card data, bank account numbers, passwords or PINs, contact information and so on. They are often the user's main store of personal information because of their portable nature. Business mobile devices often contain corporate emails and documents and as such, may contain sensitive data.

**Incidental Data Disclosure**
Data can be disclosed unintentionally by the user of a mobile device. Many users are either unaware or tend to forget that data is being transmitted or received, and most are even completely ignorant of privacy settings that can prevent this. Users are usually naïve about functionality of device applications so even though they may have given clear consent, they may not be aware that an application collects and publishes personal data(Yeboah-Boateng, Cyber-Security Challenges with SMEs in Developong Economies: Issues with Confidentiality, Integrity & Availability (CIA), 2013)(Whitman & Mattord, Pinciples of Information Security, 2011). A typical example is location information used in social networks, messages or uploaded image metadata. Inadvertent revelation of location information tends to enable attackers to track users and so allow hijacking, or robbery of vehicles having valuable merchandises. Location information is often contained within photo files and by giving an app access to the photo file, a user may be inadvertently revealing their location.

**Improper Decommissioning**
Improper decommission or the transfer of a mobile device to another user without removing sensitive data can result in an attacker gaining access to the data on it (Yeboah-Boateng, Using Fuzzy Cognitive Maps (FCMs) To Evaluate The Vulnerabilities With ICT Assets Disposal Policies, 2012). Due to a growing awareness of identity theft many people and organizations now destroy or wipe computer hard drives prior to decommissioning. Unfortunately, the same is not yet happening with mobile devices used in the workplace with sensitive corporate data. Similarly, devices being recycled are on the increase. According to a market analysts, ABI Research predicted that by 2012 over one hundred (100) million mobile devices will be salvaged every year to be re-used (Hogben & Dekker, 2010). It cannot be over-emphasized that mobile devices contain large amounts of sensitive information which may be valuable to an attacker. A study conducted by ENISA in 2010 revealed that, out of the 26 business smartphones bought second-hand on eBay, 4 contained information from which the owner could be identified while 7 contained enough data to identify the owner's employer(Hogben & Dekker, 2010). One smartphone was traced to a senior sales director of a company, and his call history, address book entries, diary, and e-mails, were recovered.

**Phishing & SMiShing**
Phishing or SMiShing can result from an attacker using phony applications, SMS, or email that appear unpretentious to collect user credentials like password, PINs, or credit card information(Yeboah-Boateng & Amanor, Phishing, SMiShing & Vishing: An Assessment of Threats against Mobile Devices, 2014). Phishing attacks are well-known threats for users of traditional computers and are increasingly becoming a concern for mobile devices and platforms alike for several reasons. For instance, the reduced screens sizes of these devices, makes it easy for attackers to camouflage useful hints like whether the website uses SSL, that users rely on to decide whether or not to submit credentials. Also, app-stores provide a new way of phishing by giving attackers the chance to place counterfeit apps in the app-store, looking like authentic apps, as well as these devices provide additional channels that can be used for phishing, using SMS in the case of SMiShing. Users may be less cautious about SMS phishing messages, and finally even though users may be aware of the risk of phishing in traditional computers, most are unaware of the same type of risk in mobile devices.

**Malware**
There are many types of malicious software (malware) and are referred to by different names depending on the function. Common malwares include Spyware, Virus, Financial Malware, Surveillance Malware, Ransomware, Trojan horse, etc.

**Network Spoofing**
A Network Spoofing attack can be the result of an attacker deploying a fake network Access Point and users unknowingly connecting to it. The attacker thereafter hijacks the user communication to carry out additional attacks such as Phishing. For mobile device users, the risk is even higher because security indicators like a 'trusted SSL connection' indicator are harder to find or missing on these devices.

**Network Congestion**
Network congestion can result from network resource overload due to many mobile devices connecting to the corporate network and depleting scarce resources and rendering them unavailable to legitimate users. Thus, the availability security dimension is said to have been compromised of breached. The uptake of smartphones usage and mobile Internet, have increased the risk of network congestion through either signaling overload or data capacity overload(Karygiannis & Owens, 2002).

## 2.4 Mitigation Measures

For each risk explained there are certain measures which could be adopted to mitigate the negative effects of realizing the risk. The following sub-sections identify these measures based on individual risks.

**Data Leakage**
To mitigate against the risk of Data Leakage, mobile devices should always be configured with an inactivity timer set together with automatic lock. This would at least make it more difficult for an attacker to gain access to the data of a lost or stolen device. Also regular backups of data on mobile devices should be made. In most cases, an automatic backup procedure would be most appropriate. This would at least ensure that sensitive and important data is retained even after the device goes missing, and encryption should be implemented for mobile device memory and removable media. This would ensure confidentiality and prevent Data Leakage on the unexpected event of a device loss(CITRIX, 2013).

**Unintentional Data Disclosure**
In order to mitigate against the risk of Unintentional Data Disclosure, extreme caution should be observed dealing with permission requests when using, or most especially, when installing device apps(Nitze, 2014). For instance, a social networking app may request access to the smartphone's address book in order to publish it on the Internet. Utmost caution should be observed with such requests. It is also recommended that default privacy settings of mobile device apps or services be reviewed and the settings changed if need be. For instance, the settings regarding whether or not to attach location information to photos, to social network posts, and so on need to be reviewed before permission is granted or denied.

**Decommissioned Devices**
Prior to decommissioning or recycling, an appropriate procedure needs to be observed. This usually would include memory wipe processes and this should apply to both removable media and on-board memory. There are standard procedures for memory wipe which ensure that data on the devices do not get leaked. Any of such procedures can be adopted, such as the NIST standard(NIST-800, 2012).

**Phishing & SMiShing Attacks**

Whereas cyber-attacks have become sophisticated, the end-users remain unsophisticated. Security education, training and awareness (SETA) should be created for this risk as most users are naïve about the fact that it could occur on mobile devices just as with traditional computers(Yeboah-Boateng & Amanor, Phishing, SMiShing & Vishing: An Assessment of Threats against Mobile Devices, 2014). A cynical attitude should be adopted to messages, content and software, especially when coming from unfamiliar sources via SMS, Bluetooth, email, and the likes.

**Malware – Spyware, Surveillance, Virus, Financial Malware etc.**

To protect against various kinds of malware, always configure mobile device to request for a PIN or password before allowing new apps to be installed. This would prevent deliberate malware installation when an attacker gains physical access to a device. This could also have the potential of preventing certain social engineering attacks. Also, resource usage on mobile devices should be closely and constantly monitored for anomalies. Anomalies would be an indication that a malware may have been installed on the device(Ernst & Young, 2012). And finally, for mobile devices with sensitive corporate information, a whitelist of apps should be defined, and app installation on those devices restricted to the list. This is because many apps are given easy read-access to the contact or address book information on devices, and information on such devices could be extremely sensitive, so it is necessary to restrict malicious apps from gaining access.

**Network Spoofing**

To protect against network spoofing, always use secure connections for corporate data communications like VPNs and SSL which ensure that communications are encrypted and always configure pre-installed public key certificates for all corporate servers, email, Intranet, etc. and configure clients to reject any certificate other than those configured(Juniper Networks, 2012).

**Network Congestion**

To minimize network congestion and address signaling overload, any mechanisms that change how frequently a mobile device switches between idle and active mode, such as the 3GPP Fast Dormancy mechanism can be adopted. With regards to data capacity as opposed to signaling overload, solutions such as LTE and WiMAX have the potential for enhancements in spectral efficiency, amount of data that can be transmitted over the air using the same amount of allocated spectrum(Juniper Networks, 2012).

### 3. METHODOLOGY

This section presents the methodology and approach used for the study. The research design used was exploratory. It started by exploring through a systematic Literature Review, the Risks, Threats, and Vulnerabilities presented by the adoption and use of BYOD, explaining their features and characteristics. It then proceeded to evaluate Mitigation measures that could be adopted to mitigate against the negative effects of the explained Risks. It further proceeded to estimate the Most Significant Risk from the taxonomy of Risks.

The matrix used to estimate the value of risk was based on that used in the European Networks and Information Security Agency (ENISA) 2010 report(Hogben & Dekker, 2010)which was derived from a product of the Likelihood and Impact of each Risk. The Likelihood and Impact of each identified Risk was determined in consultation with a group of experts. These were Information Technology and Security professionals working in various organizations. These experts were required to indicate the plausible

Likelihood, from the tuple {"Very Low" to Very High"} and Impact, {"Very Low" to Very High"} of each Risk. The average value of each Risk is reported in section 4. The model is summarized in the diagram below.

| Impact \ Likelihood | Very Low = 1 | Low = 2 | Medium = 3 | High = 4 | Very High = 5 |
|---|---|---|---|---|---|
| Very Low = 1 | 1 | 2 | 3 | 4 | 5 |
| Low = 2 | 2 | 4 | 6 | 8 | 10 |
| Medium = 3 | 3 | 6 | 9 | 12 | 15 |
| High = 4 | 4 | 8 | 12 | 16 | 20 |
| Very High = 5 | 5 | 10 | 15 | 20 | 25 |

**Figure 3-1: Schematic for Risk determination (Source:** (Hogben & Dekker, 2010)**)**

The value of the Impact and Likelihood were assigned values from "1 to 5", depending on their perceived severity. The Risk is considered a product of Likelihood and Impact, therefore to estimate the Risk, given the value of the Impact and that of Likelihood, the product of the assigned values are calculated. A risk value of "1 – 4" is considered "Low". A value of "5 – 14" is considered "Medium", and "15 – 25" is considered "High".

**Population and Sampling**
The population for this research consisted of IT functionaries and Security professionals, who manage various enterprise networks. It was perceived that these professionals possess the requisite knowledge to provide the required information to get the most accurate results from the study.

The sampling method employed was expert sampling which is a purposive sampling method. Purposive sampling considers the judgement of the researcher when it comes to selecting the units like people, cases or organizations, events, pieces of data and so on that are to be studied. Seventy-five (75) ICT functionaries and security professionals were targeted via emails, but only fifty-six (56) responses were received, which forms the basis of analysis in this study.

4.  **RESULTS AND ANALYSIS**
This section presents the results and analysis from the study.

**4.1    Characteristics of BYOD use**
This sub-section analyses the data on the characteristics of BYOD.

**Perception on Risk to Information Security posed by the adoption of BYOD**
The chart below depicts the information gathered from respondent's perception on risk to Information Security posed by BYOD adoption.

Five (5) respondents, representing 8.9% were of the view that it was not a problem at all. Two (2) of the respondents, representing 3.6% thought it was not much of a problem. Twenty-four (24) respondents, representing 42.9% towed a middle line and agreed that it was somewhat of a problem, twelve (12) representing 21.4% said it was definitely a problem, while thirteen (13) representing 23.2% affirmed that it was definitely a problem.

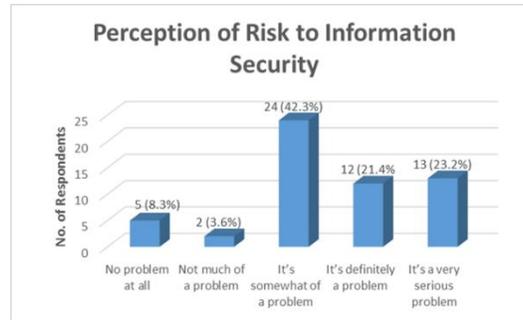

**Figure 4-1Perception on risk to Information Security posed by the adoption of BYOD (Source: Field Work, August 2015)**

The general information therefore gathered from the response to this question is that respondents agree that BYOD poses risks to information security. However the distribution clearly tells us that there is still some measure of ambiguity as opposed to being definite about the associated risks. This gives an impression that the issues concerning risks associated with BYOD use are not clearly understood and appreciated.

### 4.2  Company information accessed by BYODs on the corporate network

The figure below depicts the information gathered from respondent's type of corporate information accessed by BYODs.

Fifteen (15) respondents, representing 26.8% accessed email only. Another fifteen (15) of the respondents, representing 26.8% accessed email and calendar & scheduling applications. All other responses with smaller frequencies accessed emails together with other applications. Eight (8) however accessed none.

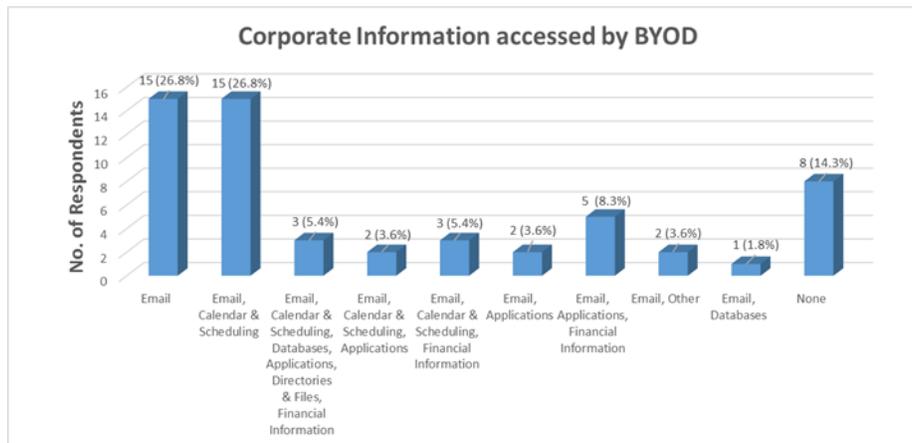

**Figure** Error! No text of specified style in document.**-2Corporate Information accessed by BYOD**

**(Source: Field Work, August 2015)**

This clearly shows that the most accessed application by BYOD was email, followed by calendar and scheduling applications. This affirms the risk exposure to corporate information since sensitive corporate emails most often reside on employees BYODs, therefore an unprotected device, not supported by IT poses the greatest danger to corporate information assets(Moody & Walsh, 1999).

**Types of Security Measures used on Mobile Devices**
The figure below depicts the information gathered from respondent's type of security measures used on employee's devices.

Sixteen (16) respondents, representing 28.6% said they used passwords or PINs only to secure their devices. Thirteen (13) of the respondents, representing 23.2% used passwords or PINs together with automatic locks, three (3) respondents, representing 5.4% used passwords or PIN, automatic locks and anti-malware, five (5) respondents, representing 8.9% used passwords or PIN, automatic locks, anti-malware, remote wipe, and VPN, two (2) respondents, representing 3.6% used passwords or PIN, automatic locks and remote wipe, two (2) respondents, representing 3.6% used passwords or PIN, automatic locks, remote wipe, and VPN, seven (7) representing 12.5% used automatic lock only, three (3) representing 5.4% used remote wipe only, two (2) representing 3.6% used VPN only, while two (2) representing 3.6% used no security measures at all.

The inference gathered from the responses is that passwords and PINs are the most used security measures, followed closely by automatic locks, since most of the other security measures are used in conjunction with passwords or PINs, and automatic locks. It is striking to note that a very important and easily available security measure like anti-malware and remote wipe which enable you to remotely wipe all data on a lost device has a very low use rate. This clearly highlights ignorance in security awareness and practices amongst employees, which in turn increases the risk exposure of corporate information assets(Jeon, Kim, Lee, & Won, 2011).

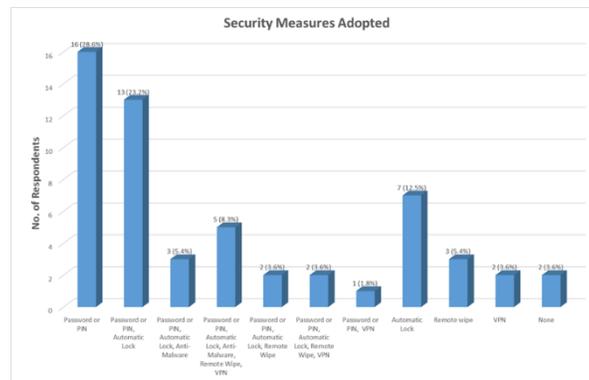

**Figure** Error! No text of specified style in document.**-3: Security measures used on BYOD (Source: Field Work, August 2015)**

**4.4     BYOD Policy deployed in organizations**
The figure below depicts the information gathered from respondent's BYOD policy deployed in organization.

Nine (9) respondents, representing 16.1% affirmed that BYODs were allowed and subsidized in the organization. Fifteen (15) of the respondents, representing 26.8% affirmed that BYODs were allowed, but not subsidized. Eighteen (18) representing 32.1% confirmed that BYODs were completely prohibited in the organization, while fourteen (14) representing 25% said there was no policy regarding BYODs at all.

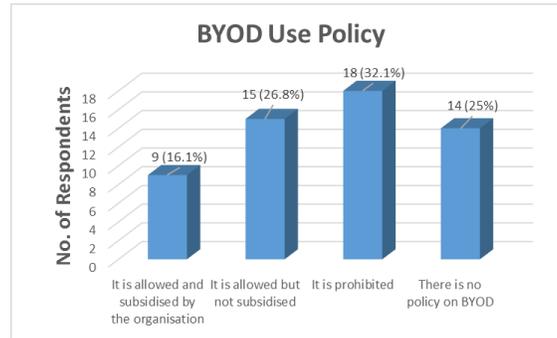

**Figure** Error! No text of specified style in document.**-4: BYOD use policy (Source: Field Work, August 2105)**

An interesting inference can be drawn from the responses received from this subsection, and the responses from subsection 4.2.2 on Corporate Information accessed by BYOD. In this subsection, 32.1% affirmed that BYODs were prohibited, whereas 25% claimed there was no policy, implying that 57.1% of respondents did not have support for BYOD. However in the above section 4.2.2 only 14.3% of respondents indicated that they did not access any corporate information at all. It is therefore clear that some employees whose organizations do not have support for BYODs still use them on the corporate network, which tend to increase the risk exposure to information security.

### 4.5    Risk Analysis
This section analyses the data on the Risk Analysis section of our questionnaire.

**Data Leakage**
The table and chart below depict responses received regarding respondents perceived risk from Data Leakage.

 Four (4) of the respondents representing 7.1% perceived the risk as "Low". Ten (10) representing 17.9% perceived the risk as "Medium", while forty-two (42) representing 75% perceived risk as "High".

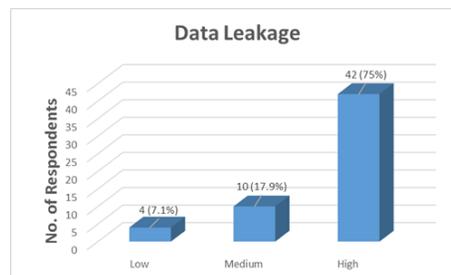

**Figure** Error! No text of specified style in document.**-5: Risk from data leakage (Source: Field Work, August 2015)**

Results show that respondents clearly appreciate the risk to Information Security posed by data leakage, and the majority of them perceived this risk as high. This is not too surprising as the features of BYODs being portable and small is size make it susceptible to theft or loss, which invariably results in sensitive data, if not properly secured, falling into wrong hands.

**Attacks on decommissioned mobile devices**

The figure below depict responses received regarding respondents perceived risk from attacks on decommissioned mobile devices or smartphones.

Six (6) of the respondents representing 10.7% perceived the risk as "Low". Twenty-three (23) representing 41.1% perceived the risk as "Medium", while twenty-seven (27) representing 48.2% perceived risk as "High".

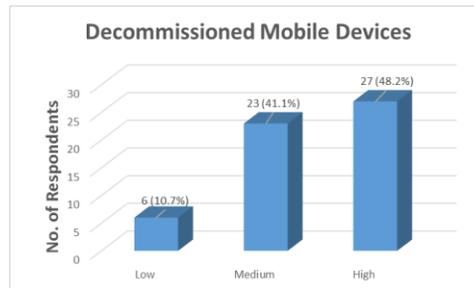

**Figure** Error! No text of specified style in document.**-6 Risks from attacks on decommissioned devices**

**(Source: Field Work, August 2015)**

Results from this survey was also somewhat clear that the risk posed by decommissioned devices was high, however a good number of respondents also perceived the risk as medium most likely because most BYODs were privately owned and therefore did not follow strict policies for decommissioning.

**Unintentional disclosure of data**

The figure below depicts responses received regarding respondents perceived risk from unintentional disclosure of data.

Four (4) of the respondents representing 7.1% perceived the risk as "Low". Twenty-five (25) representing 44.6% perceived the risk as "Medium", while twenty-seven (27) representing 48.2% perceived risk as "High".

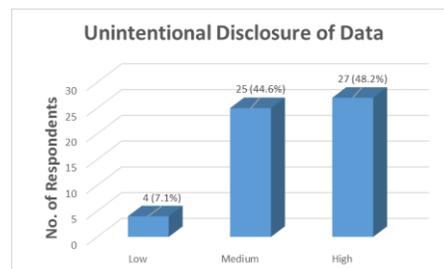

**Figure** Error! No text of specified style in document.**-7: Risk from unintentional disclosure of data**

**(Source: Field Work, August 2015)**

The results from the survey on the risks posed by unintentional disclosure of data is similar to that gotten from decommissioned devices for good reason, as these two risks are related because attacks on improperly decommissioned devices would usually result in unintentional disclosure of data.

**Phishing or SMiShing**

The figure below depicts responses received regarding respondents perceived risk from Phishing or SMiShing. Four (4) of the respondents representing 7.1% perceived the risk as "Low". Twenty-three (23) representing 41.1% perceived the risk as "Medium", while twenty-nine (29) representing 51.8% perceived risk as "High".

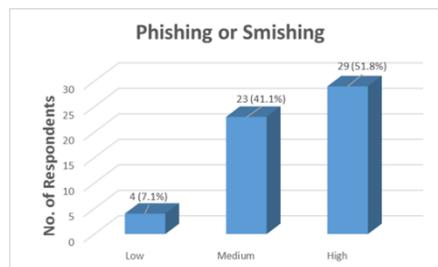

**Figure** Error! No text of specified style in document.**-8: Risk from Phishing or SMiShing (Source: Field**

**Work, August, 2015)**

Majority of the respondents for the survey on phishing and SMiShing perceived the risk as high even though a sizable percentage also perceived it as medium. This is a clear indication that the risk posed by phishing or SMiShing in BYODs is quite well appreciated, even though it was previously considered a threat in traditional computers and laptops.

**Spyware**

The figure below depicts responses received regarding respondents perceived risk from Spyware. Nine (9) of the respondents representing 16.1% perceived the risk as "Low". Nineteen (19) representing 33.9% perceived the risk as "Medium", while twenty-eight (28) representing 50% perceived risk as "High".

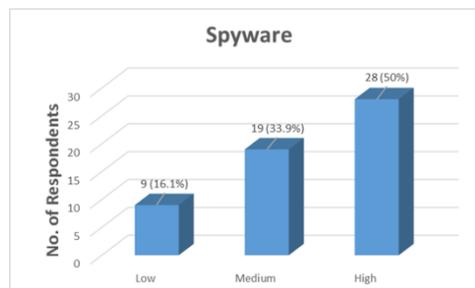

**Figure** Error! No text of specified style in document.**-9: Risk from Spyware (Source: Field Work, August 2015)**

The risk posed by spyware was generally perceived as high and indicated by 50% of respondents. Being classified a malware, spyware are not the most commonly encountered malware as the negative effects usually go unnoticed, and this is reflective the other 50% of the respondents indicating a perceived risk of medium and low.

**Financial Malware**
The figure below depicts responses received regarding respondents perceived risk from Financial Malware. Seven (7) of the respondents representing 12.5% perceived the risk as "Low". Twenty-eight (28) representing 50% perceived the risk as "Medium", while twenty-two (21) representing 37.5% perceived risk as "High".

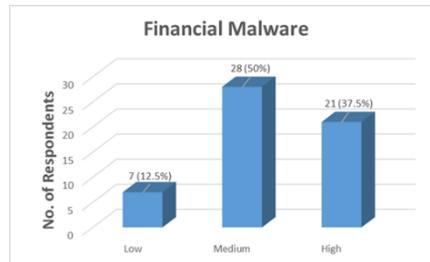

**Figure** Error! No text of specified style in document.**-10: Risk from Financial Malware (Source: Field Work, August 2015)**

Again, in the survey for financial malware, 50% of respondents perceived the risk as medium. This can be attributed to the fact that the target of this type of malware is financial information; however there is generally a limited use of BYODs and smartphones for financial transactions amongst the surveyed population, as the general adoption of online transactions has been rather low due to skeptics on security concerns.

**Viruses**
The figure below depicts responses received regarding respondents perceived risk from Viruses. Four (4) of the respondents representing 7.1% perceived the risk as "Low". Twelve (12) representing 21.4% perceived the risk as "Medium", while forty (40) representing 71.4% perceived risk as "High".

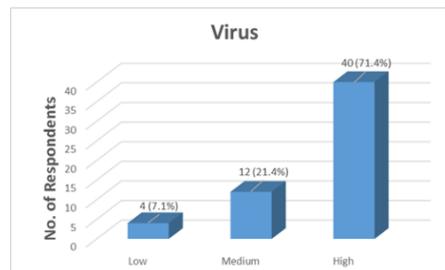

**Figure** Error! No text of specified style in document.**: Risk from Virus (Source: Field Work, August 2015)**

Viruses however, are the most commonly encountered malware, and this is indicative of a high percentage of respondents indicating a perception of high risk.

**Network Congestion**
The figure below depicts responses received regarding respondents perceived risk from Network Congestion. Fourteen (14) of the respondents representing 25% perceived the risk as "Low". Sixteen (16) representing 28.6% perceived the risk as "Medium", while twenty-six (26) representing 46.4% perceived risk as "High".

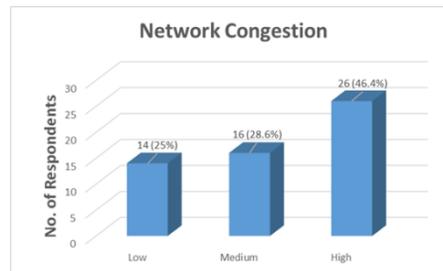

**Figure** Error! No text of specified style in document.**: Risk from Network Congestion (Source: Field Work, August 2015)**

The perception of risk from network congestion from the responses is quite neutral and therefore can be considered that respondents generally do not perceive network congestion as much of a problem.

**Network Spoofing**
The figure below depicts responses received regarding respondents perceived risk from Network Spoofing. Nine (9) of the respondents representing 16.1% perceived the risk as "Low". Twenty-one (21) representing 37.5% perceived the risk as "Medium", while twenty-six (26) representing 46.4% perceived risk as "High".

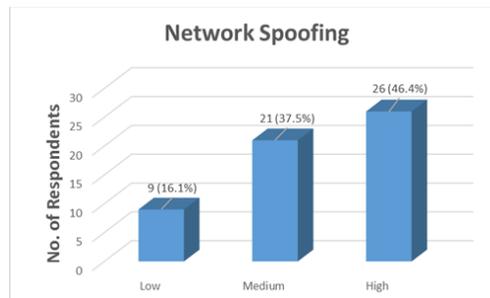

**Figure** Error! No text of specified style in document.**:.Risk from Network Spoofing (Source: Field Work, August 2015)**

The perception to the risk associated with network spoofing is pretty much similar to that of network congestion, as seen from the responses and respondents generally do not perceive that as much of a problem.

**Relative Importance Index of Risk**

The figure below depicts the relative importance index of the various risks.

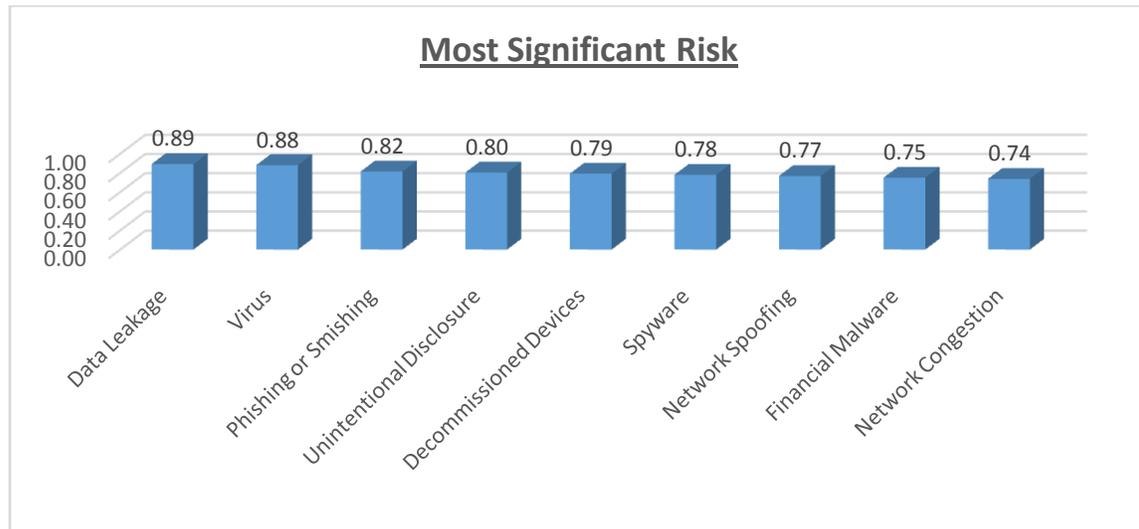

**Figure** Error! No text of specified style in document.**14: Relative Importance Index of identified risks**

**(Source: Field Work, August 2015)**

The most significant risk was evaluated from all the responses from the analysis of the various risks. Data Leakage was found to be the highest, followed closely by Virus, which is very common as it has been inherited from traditional computers. The reasons for Data Leakage being perceived as the most significant risk is that, its portable nature exposes it to theft or loss, and if data on these devices are not properly protected, its loss can invariably result in data leakage.

5. **CONCLUSION**

This section discusses the major findings, the implications for organisations, and makes some recommendations based on the findings.

**Study Findings**

There were several interesting observations and findings gained from this research.

Firstly, it was observed from respondent's perception of BYOD use that it posed cybersecurity challenges to corporate information assets.

Secondly, a lot more people were using BYODs on the corporate networks without the knowledge of management, and there were no specific policies covering BYOD use in many organisations which posed an increased security threat to corporate information assets. Particularly, in response to the question "What kind of corporate information do you access with your BYOD?" only eight (8) representing 14.3% responded "None", implying that forty-eight (48) representing 85.7% did use BYODs on the corporate network. However, in contrast, in response to the question "What is the BYOD use policy in your organisation?" eighteen (18) representing 32.1% responded that it was prohibited, while fourteen (14) representing 25% responded that there was no policy at all. This confirms that in most organisations,

BYOD use is not being monitored and controlled appropriately in order to effectively and successfully mitigate against the risky impacts.

Thirdly, it was determined that the most significant risk identified was that of data leakage which was in consonance with the ENISA 2010 report on a study into smartphone security risks.

Fourthly, many people do not use security measures available to mitigate against risks. This can be deduced from responses to the question, "What security measures do you use on your BYOD?", where it was obvious that the "Remote Wipe" feature that could remotely control a lost device had a very low rate of use, even though the most significant risk identified was found to be data leakage.

In conclusion, the phenomenon of BYOD is here to stay and in the near future, as the cost of BYODs decrease and therefore become more available, the number of BYODs connecting to the corporate network is bound to increase and therefore there is the need for management to embrace the development by understanding the risks, leveraging the benefits, and adoptthe necessary security measures to mitigate against the risky effects.

**Recommendations**
From the research findings, the researcher recommends that management needs to keep close track of the BYOD phenomenon, in order to control its development to achieve a positive rather that a negative effect on business.

Areas that need to be addressed are as follows:

1. Enhance employee security awareness
2. BYOD use policy adoption
3. Monitoring and control of BYOD use

**Further Research**
The researcher recommends that further studies be conducted into evaluating the impact of cybersecurity breaches on the business. This would give management a clear picture on what they stand to lose thereby helping to encourage investments into appropriate mitigation strategies.

**References**

Aittokallio, A. (2014). *Mobile Malware Infection Rate Accelerating.* Telecoms.Com.

Belletati, G. (2014). *Building Digital Workplaces: The BYOD Approach.* Universita Ca' Fascan Venezia.

Causey, B. (2013, January 1). Strategy: How to Conduct an Effective IT Security Risk Assessment. *Information Week*. Retrieved from http://reports.informationweek.com/abstract/21/9796/Security/Strategy:-How-to-Conduct-an-Effective-IT-Security-Risk-Assessment.html

CEBR. (2015). *The Business and Economic Consequences of Inadequate Cybersecurity.* Center for Economic & Business Reserach (CEBR).

CITRIX. (2013). *Best Practices to make BYOD simple and secure: A Guide to Selecting Technologies and Developing Policies for BYOD.* Citrix Systems, Inc.

Ernst & Young. (2012). *Mobile Device Security - Understanding Vulnerabilities & Managing Risks.* Ernst & Young (EY).

Ernst & Young. (2013). *Security & Risk Considerations for your Mobile Device Program.* Ernst & Young (EY).

Field, T. (2011). *BYOD: Manage the Risks*. Retrieved August 20, 2015, from http://www.bankinfosecurity.co.uk/interviews/byod-manage-risks-i-1327

Ghosh, A., Gajar, P. K., & Rai, S. (2013). Bring-Your-Own-Device (BYOD): Security Risks and Mitigating Strategies. *Journal of Global Research in Computer Science, 4*(4).

Hogben, G., & Dekker, M. (2010). *Smartphone Security: Information Security Risks, Opportunities and Recommendations for Users.* London: ENISA.

Jeon, W., Kim, J., Lee, Y., & Won, D. (2011). *A Practical Analysis of Smartphone Security.* Howne: Department of Cyber Investigation Police.

Juniper Networks. (2012). *2011 Mobile Threats Report*. Retrieved 08 21, 2015, from https://www.juniper.net/us/en/local/pdf/additional-resources/jnpr-2011-mobile-threats-report.pdf

Karygiannis, T., & Owens, L. (2002). *Wireless Network Security for 802.11, Bluetooth, and Handheld Devices.* National Institute of Standards and Technology (NIST).

Moody, D., & Walsh, P. (1999). Measuring The Value of Information: An Asset Valuation Approach. *European Conference on Information Systems (ECIS 99)* (pp. 1-17). Melbourne: European Conference on Information Systems (ECIS 99).

NIST-800. (2012). *Guide for Conducting Risk Assessements.* National Institute of Standards & Technology.

Nitze, A. (2014). The Case for Web Technologies in Mobile Business Apps. *Web Technologies for Business Apps*, 87-96.

Nunoo, E. M. (2013). Smartphone Information Security Risks. Stockholm: Lulea University of Technology.

Whitman, M. E., & Mattord, H. J. (2004). *Management of Information Security* (illustrated ed.). Course Technology.

Whitman, M. E., & Mattord, H. J. (2011). *Pinciples of Information Security* (4th ed.).

Yeboah-Boateng, E. O. (2012, October). Using Fuzzy Cognitive Maps (FCMs) To Evaluate The Vulnerabilities With ICT Assets Disposal Policies. *International Journal of Electrical & Computer Sciences IJECS-IJENS, 12*(5), 20-31.

Yeboah-Boateng, E. O. (2013). Cyber-Security Challenges with SMEs in Developong Economies: Issues with Confidentiality, Integrity & Availability (CIA). Copenhagen: Institut for Elektroniske Systemer, Aalborg Universitet.

Yeboah-Boateng, E. O., & Amanor, P. M. (2014, April). Phishing, SMiShing & Vishing: An Assessment of Threats against Mobile Devices. *Journal of Emerging Trends in Computing and Information Sciences, 5*(4), 297-307.